\documentclass[english,10pt,aps,prb,twocolumn,notitlepage,superscriptaddress] {revtex4-1}

\usepackage{graphicx}
\usepackage{amssymb}
\usepackage{amsmath}
\usepackage{setspace}
\usepackage{babel}
\usepackage{epstopdf}
\usepackage{color}

\definecolor{OliveGreen}{rgb}{0,0.6,0}

\usepackage{lineno}

\makeatletter

\begin{document}
\title{Pauli spin blockade in CMOS double quantum dot devices}

\author{D. Kotekar-Patil}
\email[]{dharamkotekar@gmail.com}
\affiliation{Universit\'e Grenoble Alpes, INAC-PHELIQS, F-38000 Grenoble, France}
\affiliation{INAC-PHELIQS, CEA Grenoble, F-38000 Grenoble, France}
\author{A. Corna}
\affiliation{Universit\'e Grenoble Alpes, INAC-PHELIQS, F-38000 Grenoble, France}
\affiliation{INAC-PHELIQS, CEA Grenoble, F-38000 Grenoble, France}
\author{R. Maurand}
\affiliation{Universit\'e Grenoble Alpes, INAC-PHELIQS, F-38000 Grenoble, France}
\affiliation{INAC-PHELIQS, CEA Grenoble, F-38000 Grenoble, France}
\author{A. Crippa}
\affiliation{Universit\'e Grenoble Alpes, INAC-PHELIQS, F-38000 Grenoble, France}
\affiliation{INAC-PHELIQS, CEA Grenoble, F-38000 Grenoble, France}
\author{A. Orlov}
\affiliation{Department of Electrical Engineering, University of Notre Dame, USA}
\author{S.Barraud}
\affiliation{Universit\'e Grenoble Alpes,  F-38000 Grenoble, France}
\affiliation{LETI MINATEC campus, CEA Grenoble, F-38054 Grenoble, France}
\author{L. Hutin}
\affiliation{Universit\'e Grenoble Alpes,  F-38000 Grenoble, France}
\affiliation{LETI MINATEC campus, CEA Grenoble, F-38054 Grenoble, France}
\author{M. Vinet}
\affiliation{Universit\'e Grenoble Alpes,  F-38000 Grenoble, France}
\affiliation{LETI MINATEC campus, CEA Grenoble, F-38054 Grenoble, France}
\author{X. Jehl}
\email[]{xavier.jehl@cea.fr}
\affiliation{Universit\'e Grenoble Alpes, INAC-PHELIQS, F-38000 Grenoble, France}
\affiliation{INAC-PHELIQS, CEA Grenoble, F-38000 Grenoble, France}
\author{S. De Franceschi}
\affiliation{Universit\'e Grenoble Alpes, INAC-PHELIQS, F-38000 Grenoble, France}
\affiliation{INAC-PHELIQS, CEA Grenoble, F-38000 Grenoble, France}
\author{M. Sanquer}
\affiliation{Universit\'e Grenoble Alpes, INAC-PHELIQS, F-38000 Grenoble, France}
\affiliation{INAC-PHELIQS, CEA Grenoble, F-38000 Grenoble, France}

\begin{abstract}
Silicon quantum dots are attractive candidates for the development of scalable, spin-based qubits. Pauli spin blockade in double quantum dots provides an efficient, temperature independent mechanism for qubit readout. Here we report on transport experiments in  double gate nanowire transistors issued from a CMOS process on 300\,mm silicon-on-insulator wafers. At low temperature the devices behave as two few-electron quantum dots in series. We observe signatures of Pauli spin blockade with a singlet-triplet splitting ranging from 0.3 to 1.3\,meV. Magneto-transport measurements show that transitions which conserve spin are shown to be magnetic-field independent up to $B = 6$ T.
\end{abstract}

\maketitle

Recent breakthroughs in silicon spin quantum bits \cite{JJPla2012,Muhonen2014,Veldhorst2014,Veldhorst2015} justify the need to develop a truly CMOS-compatible route towards scalable, integrated qubits. Indeed the microelectronics industry routinely fabricates devices with critical dimensions well below the 100\,nm range~\cite{Natarajan2014,Barraud2012a}, which is enough to observe quantum effects at low temperature~\cite{Dupont-Ferrier2013,Voisin2016,Heorhii2016}. Our approach consists in starting with a state-of-the-art advanced microelectronics process in order to make further integration straightforward. Here we illustrate a first step towards this goal enabled by silicon-on-insulator (SOI) n-type nanowire transistors featuring two closely spaced gates parallel to each other. This layout results in two coupled quantum dots (QDs) in series, where each QD can be operated down to the few-electron regime by adjusting the voltages applied to the corresponding top gate and to the global back gate provided by the silicon substrate \cite{Roche2012}. 
We observe signatures of Pauli spin blockade (PSB) in different electronic configurations and for different values of the singlet-triplet splitting.  The PSB regime is studied as a function of external magnetic field, $B$, applied perpendicular to the chip plane. 
Since PSB is a commonly used mechanism for spin-qubit readout, our work bares relevance to the realization of CMOS-based qubits \cite{Maurand2016}.

The studied double QD devices were fabricated on
300-mm SOI wafers using an industry-standard process flow~\cite{Barraud2012a}.
This results in high device yield and quite reproducible electronic properties.
In particular, key properties such as single-electron tunneling due to
prominent Coulomb blockade effect and sizable size quantization have already been
demonstrated in single-gate devices at low temperature~\cite{RochePRL2012}. 
In order to create tunable double QDs, two closely spaced gates are required. Devices with two lateral gates facing each other and partially covering the silicon nanowire channel have allowed us to realize coupled atom transistors where transport occurs by sequential tunneling across two donors in the low-doped channel region ~\cite{Dupont-Ferrier2013, RochePRL2012}. An alternative approach is two parallel gates in a series geometry. Recently, this type of geometry was studied in the case of short gate spacers and a silicided (i.e. metallic) channel between the gates. It was shown that such devices can be operated as quantized current sources with the two top gates providing tunable tunnel barriers ~\cite{jehl2013}. 
Here we address the case of long gate spacers, preventing the silicidation of the channel region between the gates. We use the top gates to accumulate small puddles of electrons, forming few-electron QDs.  

\begin{figure}
\includegraphics[width=1\columnwidth]{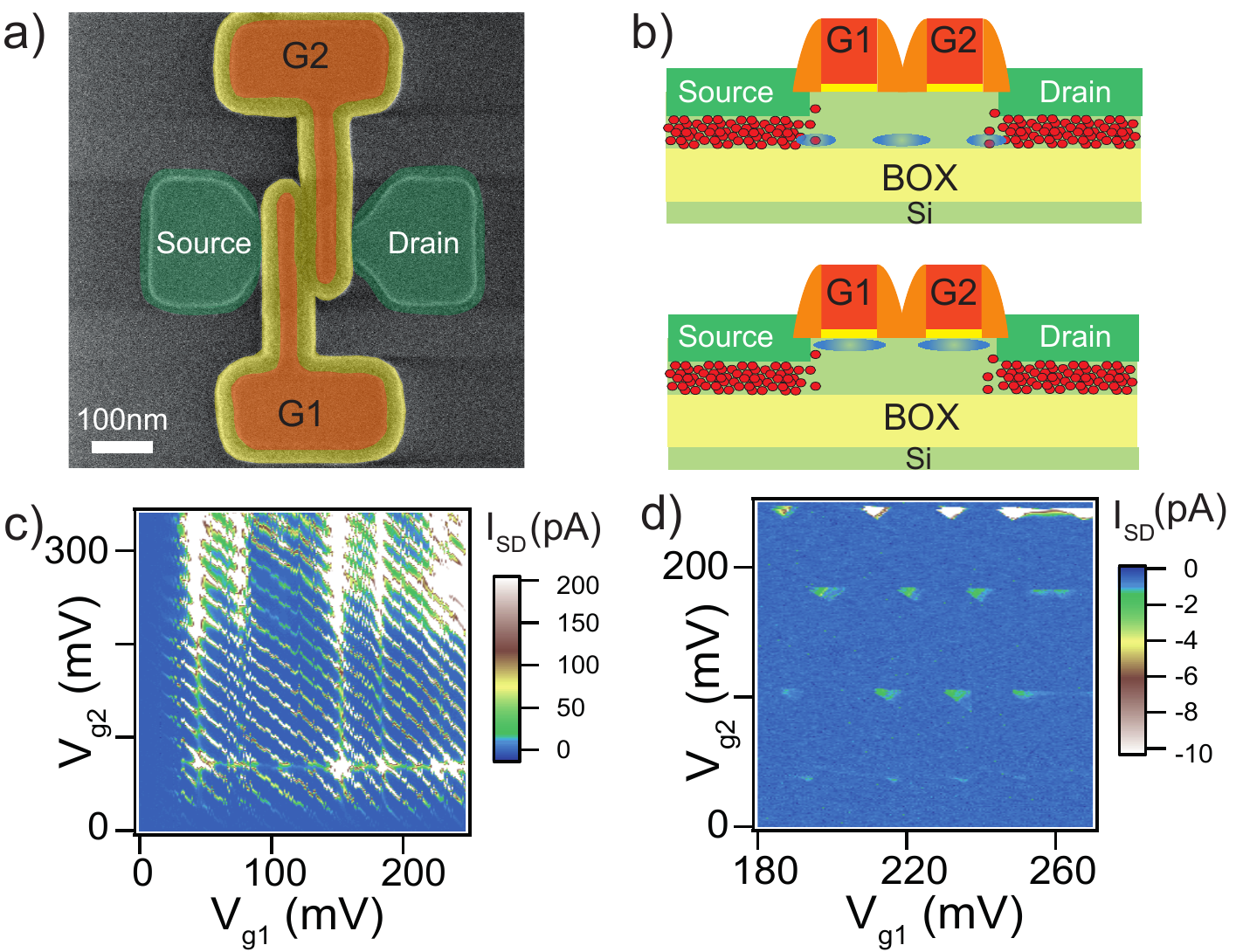}
\caption{a) False color scanning electron micrograph of a two-gate device with the source/drain highlighted in green, overlapping spacers in yellow and gates in orange. The thin nanowire is covered by the two gates and spacers in series. b) Schematic cross sections. On top, the behavior at large positive back-gate voltage ($V_{bg}$) where access resistances are lowered and a single dot is formed by depleting a 2D electron gas at the bottom interface near the BOX with the top gates. At the bottom, near zero or for lower back-gate voltage two coupled dots are formed by accumulation with the top gates. c) 2D-plot of the dc source-drain current versus both top gate voltages, recorded with a bias voltage 1\,mV at 50\,mK and $V_{bg}$= 40\,V. Anti-diagonal lines are observed, typical of a single dot controlled nearly equally by two gates. d) same plot but for $V_{bg}$=15\,V and with a bias voltage of 2\,mV; in this case triangles characteristic of two dots in series are observed.}
\label{fig1} 
\end{figure}

A scanning electron micrograph of a device and a schematic cross-sectional view are shown in Figs. \ref{fig1}a and \ref{fig1}b, respectively. The silicon channel, with a thickness of 11\,nm and a width $W$=15\,nm, is defined by deep ultra-violet lithography followed by an oxidation-etching trimming process. A 145\,nm-thick SiO$_2$ buried oxide (BOX) separates the nanowire channel from the silicon substrate. The latter is used as a back gate in order to tune the conductance of the access regions below the spacers~\cite{Roche2012}. 
The two top gates, labeled as $G1$ and $G2$, are obtained through two lithographic steps. The first one, based on conventional optical lithography, defines a large single gate. The second one, based on e-beam lithography, splits the defined gate into two 30-nm-wide lines spaced by approximately 35 nm. Such a small spacing is required to enable sufficient inter-dot tunnel coupling. 

Depending on the applied gate voltages the device can be operated in either single or double QD regime.
A single QD is obtained when a relatively large positive voltage $V_{bg}$ is applied to the silicon substrate. This creates a conducting channel near the bottom interface of the Si nanowire \cite{Khalafalla2007}. This channel can be locally depleted by the two top gates, resulting in the formation of tunnel barriers confining a single QD in the region between them (see upper diagram in Fig.~\ref{fig1}b). This transport regime is reported in Fig. ~\ref{fig1}c, where we show a measurement of dc source-drain current, $I_{sd}$, as a function of gate voltages $V_{g1}$ and $V_{g2}$ applied to $G1$ and $G2$, respectively. This data set was obtained with a constant source-drain bias voltage $V_{sd}=1$ mV.  

A double QD can instead be formed at lower values of $V_{bg}$, below the threshold for the creation of a conducting channel. In this regime, the two top gates can be used to induce charge accumulation underneath them resulting in a pair of QDs as depicted in the bottom panel of Fig.~\ref{fig1}b. A representative electron transport measurement in this regime is shown in Fig. ~\ref{fig1}d for $V_{sd} = - 2$ mV. Current conduction occurs at isolated spots corresponding to the condition $\mu_s \ge \mu_1 \ge \mu_2 \ge \mu_d$ (or, equivalently,  $\mu_d \ge \mu_2 \ge \mu_1 \ge \mu_s$ for $V_{sd} > 0$), where $\mu_s$, $\mu_1$, $\mu_2$, and $\mu_d$, are the electrochemical potentials of source reservoir, QD1, QD2, and drain reservoir, respectively (in this notation
$\mu_s - \mu_d = e V_{sd}$, where $e$ is the electron charge). 
The observed current spots have a characteristic triangular shape. Their position in the ($V_{g1}$,$V_{g2}$) plane identifies a charge boundary between consecutive occupation numbers in both QD1 and QD2. Their current intensity decreases towards the lower-left corner of Fig. ~\ref{fig1}d, where both QDs have the lowest occupation numbers and the lowest tunnel couplings to the leads and between them. Although we cannot tell the precise number of electrons on each dot we believe to have reached the few-electron occupation (see supplementary information, fig. S1). Conservatively, we could say that, in the regime of Fig  ~\ref{fig1}d,  both QDs host less than 10 electrons each. 
In the following, we shall restrict our attention to this few-electron limit and present a detailed study of selected current triangles as a function of an applied magnetic field, $B$, perpendicular to the substrate plane.


\begin{figure}
\includegraphics[width=1\columnwidth]{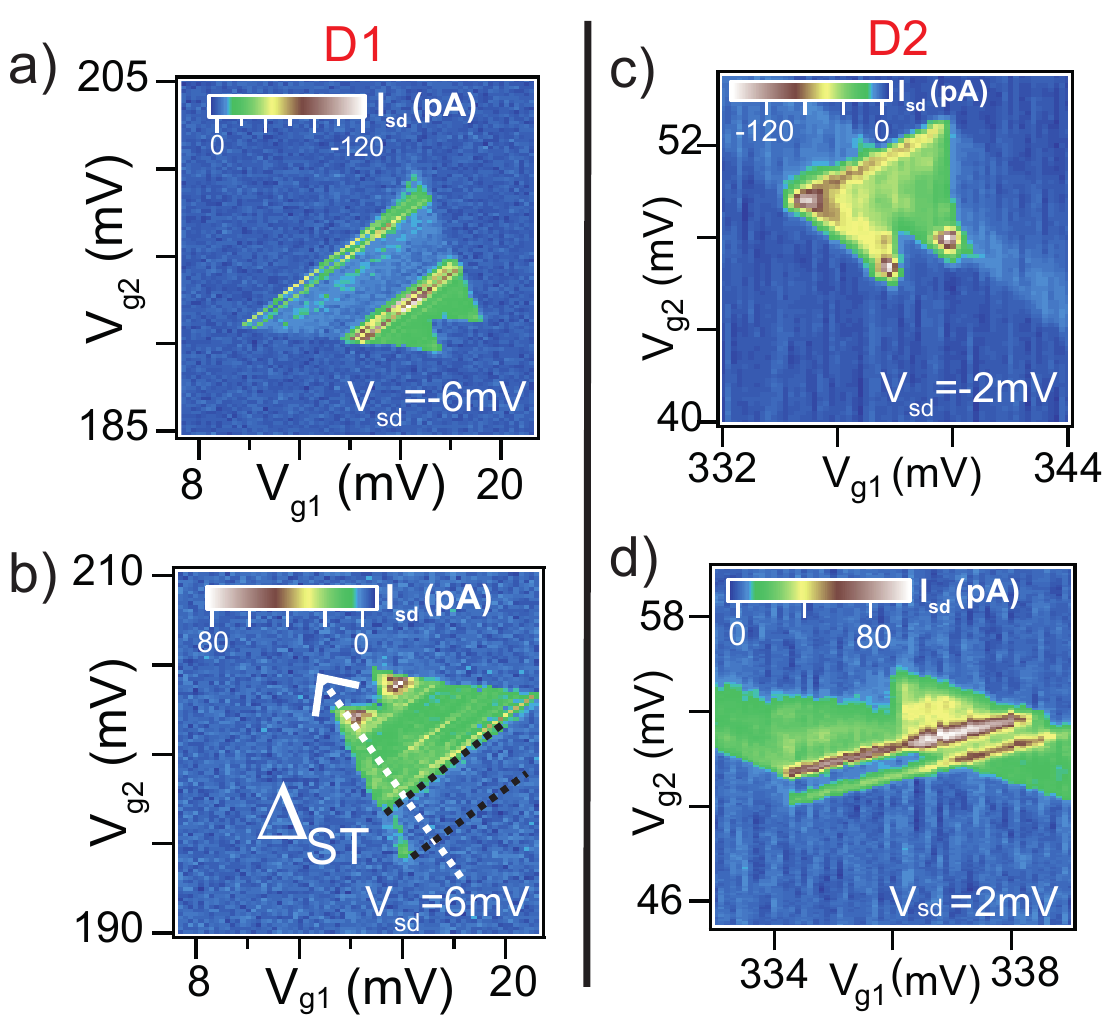}
\caption{Current $I_{\rm sd}$ as a function of top gate voltages $V_{\rm g1}$ and $V_{\rm g2}$ for two differents devices $D1$ and $D2$ respectively. Bias triangles where Pauli spin blockade is absent are presented in the upper part of the figure, respectively (a) and (c ). Their corresponding opposite polarities are presented in the lower part of the figure, respectively (b) and (d) where we identified current rectification due to Pauli spin blockade. In (b) detuning axis is indicated with a white dotted line, while the singlet triplet splitting $\Delta_{\rm ST}$ is indicated by two black dashed line. } 
\label{fig2} 
\end{figure}

Fig. ~\ref{fig2} shows $I_{\rm sd}(V_{g1},V_{g2})$ for two different devices (similar to the device geometry for which data shown in fig.\ref{fig1}c and \ref{eq1}d and operated in the few electron regime.), labelled as $D1$ and $D2$, of the type shown in Fig.~\ref{fig1}a. For device $D1$, data is  presented in Fig.~\ref{fig2}a, b and taken at a temperature of 50 mK whereas data for device $D2$ is presented in Fig.~\ref{fig2}c, d taken at a temperature of 400 mK. In each pair of data sets, the two measurements shown refer to opposite values of $V_{sd}$. 
The data reveals a suppression of $I_{sd}$ for one of the $V_{sd}$ polarities, which we attribute to PSB ~\cite{Ono2002,Johnson2005,Lai2011,Liu2008}. 
Starting with device $D1$, at $V_{\rm sd}$=-6\,mV (fig. \ref{fig2}a) we observe overlapping triangles in which no spin blockade is observed. 
Resonant transport through the ground states of the two dots is indeed visible as a current ridge at the base of each triangle, corresponding to the condition $\mu_1 = \mu_2$. Additional current ridges parallel to the bases can be identified inside the triangles. They are signatures of transport through excited states~\cite{Hanson2007}. More precisely, excited states of QD2 (QD1) if electrons tunnel from QD1 (QD2) to QD2 (QD1). The gate-voltage spacing between ground- and excited-state ridges can be translated into an energy difference of the order of a few meV, which is the typical size-quantization energy scale in these QDs \cite{House2011}. 
The background current inside the triangles is associated to inelastic tunneling between the two QDs involving phonon emission.

In the reverse polarity $V_{\rm sd}$=6\,mV (Fig.~\ref{fig2}b) we observe a truncated pair of triangles. Current suppression in the lower portion of the triangles arises from the Pauli exclusion principle preventing the transition between a spin triplet state with one electron in each dot (labelled $T_{11}$) to the singlet state with two electrons in the same dot (labelled $S_{02}$). Current is restored inside the triangles for a sufficiently large detuning, $\epsilon = \mu_1 - \mu_2$, between the chemical potentials of the two dots. 
More precisely, for a detuning value such that $T_{11}$ is aligned with the triplet state $T_{02}$ current is allowed again. Consequently, the extension of the spin blocked region gives a direct indication of the singlet-triplet splitting $\Delta_{\rm ST}$~\cite{Goswami2006}. 
It has to be emphasized that the actual number of charges in the two dots are not strictly speaking (1,1) or (0,2), but in fact $2n+1,2m+1$ and $2n,2m+2$, where $n$ and $m$ are small integers. 
As a result, assigning $\Delta_{\rm ST}$ directly to the valley-orbit splitting in our device is not appropriate here~\cite{Culcer2010}. 

Similar features are seen for device D2 shown in fig. \ref{fig2}c, d in terms of resonant tunneling through the ground and excited states of the two QDs. Additionally, we see stripes of current outside the region of the bias triangle for device D2 (fig. \ref{fig2}c, d) which are not present for device D1. We attribute these stripes of current to cotunneling of electrons through QD2 while QD1 is in Coulomb blockade.

\begin{figure}[b] 
\includegraphics[width=1\columnwidth]{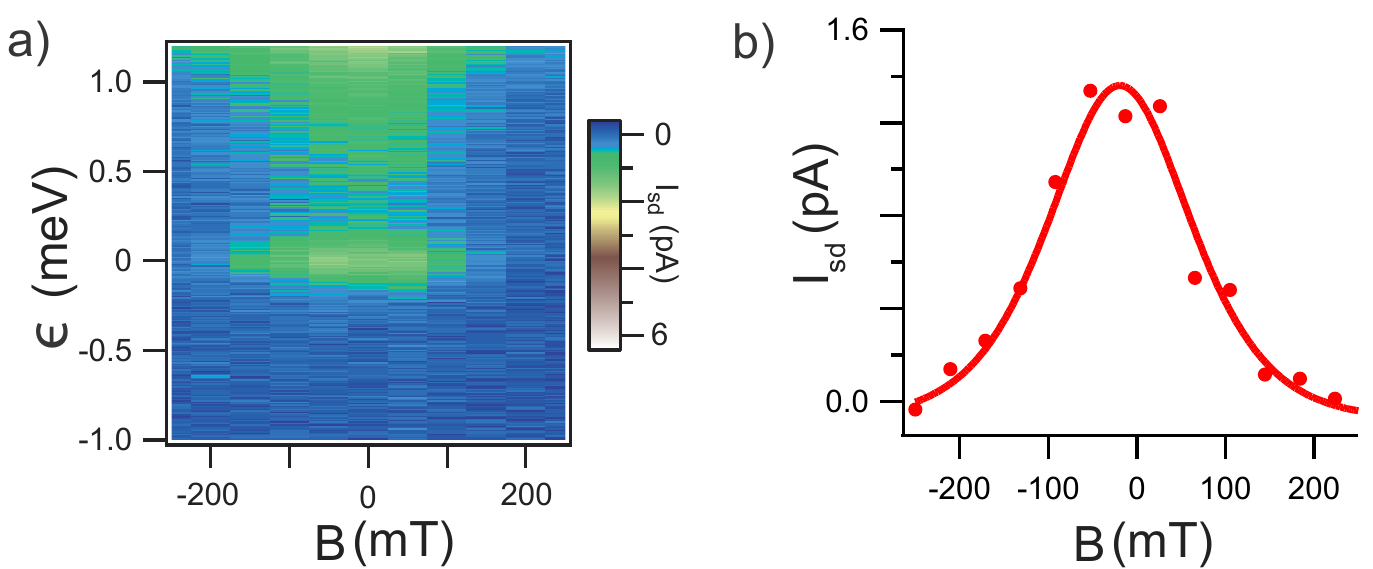} \caption{a) Leakage current as a function of magnetic field, $B$ and detuning $\epsilon$ for the data in panel \ref{fig2}b. Leakage current is suppressed for $|B|>\pm$ 200 mT. b) Linecut from data in panel a along B at $\epsilon=0$. Fit to the experimental data using equation\ref{eq1}.} 
\label{fig3} 
\end{figure}

Both the spin-blocked bias triangles presented in Fig.~\ref{fig2} exhibit a small leakage current at zero detuning i.e. when the ground states of the two dots are aligned. In Fig.~\ref{fig3}(a) we study the evolution of this leakage current versus detuning and magnetic field for the blocked region of Fig. \ref{fig2}b. For a magnetic field $|B|>$200\,mT the leakage current vanishes completely and no current can be detected for $\epsilon<\Delta_{\rm ST}$. In natural silicon the random Overhauser field is expected to be $\delta \mathcal{A}=\mathcal{A}/\sqrt{N_s}$ with $N_{s}$ the number of $^{29}$Si nuclei embedded by the electron wavefunction and $\mathcal{A}\approx$ 2\,mT is the Overhauser field for fully polarized nuclear spins~\cite{Assali2011}. Therefore hyperfine interaction cannot be at the origin of the observed leakage current. Spin-flip co-tunneling resulting in a transition from a spin-triplet state, $T_{11}$, to the spin-singlet state $S_{11}$ can provide an alternative mechanism to lift spin blockade~\cite{Qassemi2009,Coish2011}.  In that case, the magnetic-field dependence of the leakage current at zero detuning is expected to follow the relation\cite{Qassemi2009,Lai2011}:

  \begin{equation}
I_{sd}(B)=\frac{4e}{3} \Gamma^0_{\text{cotu}} \frac{g \mu_BB}{\sinh \frac{g \mu_BB}{k_B T_e}} 
  \label{eq1}
  \end{equation}  

where $\Gamma^0_{cotu}$ is the spin-flip cotunneling coupling at $B=0$, $g$ is the electron Land\'e gyromagnetic factor, $\mu_{\rm B}$ the Bohr magneton, $k_B$ the Boltzmann constant, and $T_e$ the electronic temperature in the source and drain leads.

Fig. \ref{fig3}b shows the experimental B-dependence of the leakage current at $\epsilon = 0$ as extracted from Fig. \ref{fig3}a together with a fit to equation \ref{eq1} (solid red line).  
In the fitting of \ref{fig3}b, the $g$-factor is assumed to be 2 (bare electron $g$-factor) and $\Gamma^0_{cotu}$ and $T_e$ are the fit parameters. 
The fit parameters are found to be $\Gamma^0_{cotu}= 1.03$ GHz/meV  and $T_{e} = 75$ mK, in agreement with the expected electronic temperature in the source and drain. 
Spin-flip cotunneling mechanism results in leakage current until the thermal energy of the electrons in the leads are larger than the Zeeman energy (i.e. for $3.5k_BT > g\mu_BB$).
With increasing B - field, the Zeeman energy increases. Once the Zeeman energy is larger than the thermal energy of the electrons in the leads, the leakage current drops rapidly. Using $T_e = 75$mK from the fitting in fig. \ref{fig3}b , and equating  $3.5k_BT = g\mu_BB$, we expect that the leakage current due to spin-flip cotunneling to be supressed at $B \approx 0.2$ T. This is in very good agreement with the experimental data shown in fig. \ref{fig3}b.


\begin{figure} 
\includegraphics[width=1\columnwidth]{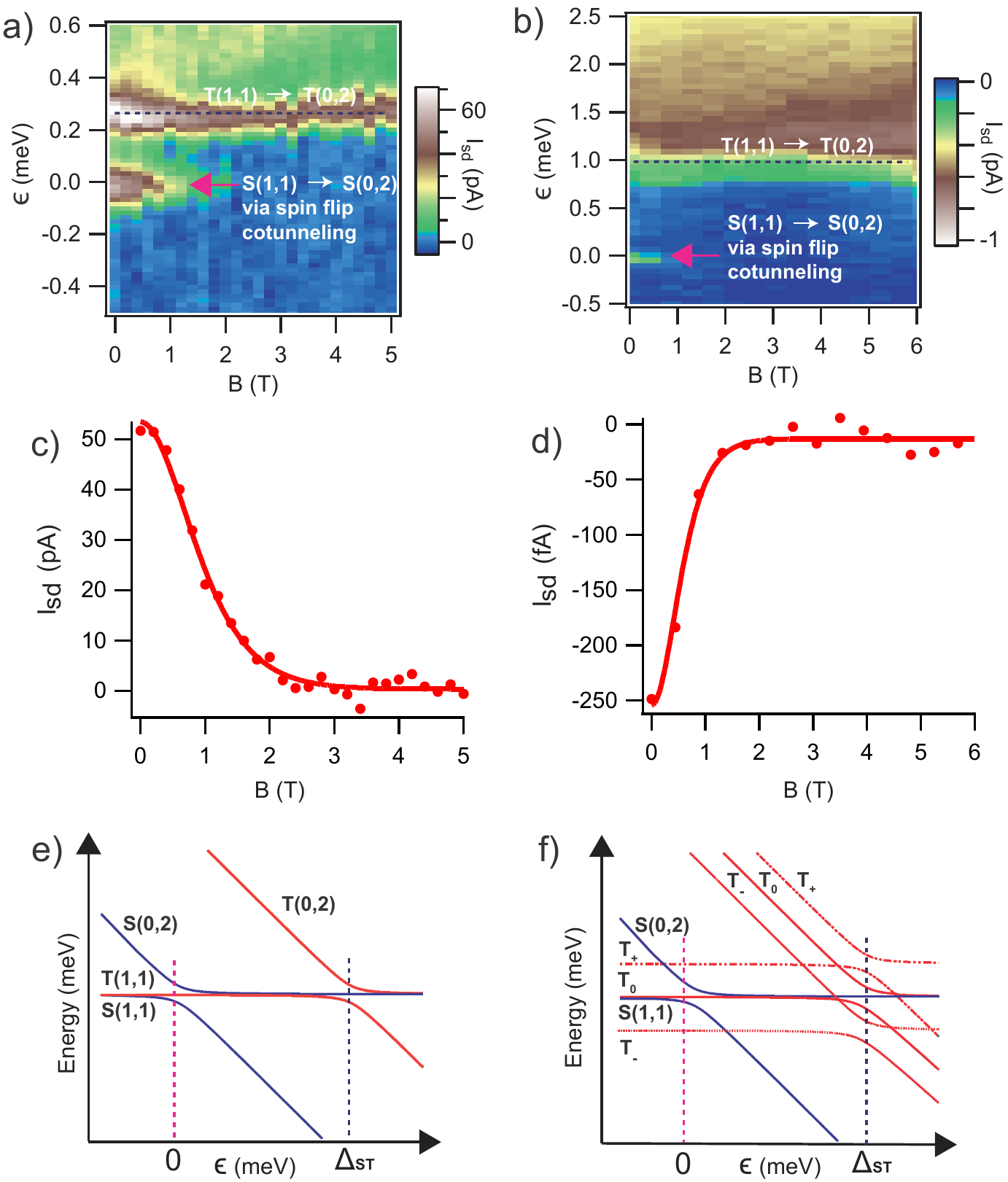} \caption{ $I_{\rm sd}$ as a function of detuning, $\epsilon$ and magnetic field, $B$. a) For the data in panel fig. \ref{fig2}d. b) for another bias triangle measured on device $D1$ (fig. S2 in supplementary information). In both the cases, leakage current is suppressed for $B>1$ T. Secondly, the first excited state, attributed to $T_{11}$ to $T_{02}$ have no magnetic field dependence. c) $I_{sd}$ vs B at $\epsilon = 0$ for fig. \ref{fig4}a. d) $I_{sd}$ vs B at $\epsilon = 0$ for fig. \ref{fig4}b.  c) Schematic of the singlet and triplet states at $B=0$ T and d) $B\neq  0$ T. } 
\label{fig4} 
\end{figure}

We now focus on the evolution of $\Delta_{\rm ST}$ for larger magnetic fields, up to $B$= 5 T. Figs. ~\ref{fig4}a and ~\ref{fig4}b show $I_{\rm sd}$ as a function $\epsilon$ and $B$ for the spin-blockade region of Fig. ~\ref{fig2}d and another bias triangle measured on device $D1$ (fig. S2 in supplementary information), respectively. These measurements were performed in different cryostats with temperature $\approx 0.4$ K (fig. \ref{fig4}a) and $\approx 0.3$ K (fig. \ref{fig4}b). At $B =0$ T and $\epsilon=0$, both data sets show a leakage current due to spin-flip cotunneling as discussed above. Because of the higher electron temperature, the suppression of the leakage current occurs on a larger $B$ range, 
$> 1$ T, as opposed to the $B \approx 0.2$ T found before (Fig. \ref{fig3}b). Fig.\ref{fig4}c and \ref{fig4}d show line-cuts at $\epsilon = 0$ along with a fit to equation \ref{eq1} (solid red line) for fig.\ref{fig4}a and \ref{fig4}b respectively. The $T_e$ and $\Gamma^0_{cotu}$ extracted from the fittings are 570 mK and 5.06 GHz/meV for fig.\ref{fig4}c and 357 mK and 36.65 MHz/meV for fig.\ref{fig4}d, respectively. The expected magnetic field to suppress the leakage current due to spin-flip cotunneling using the $T_e$ from fittings in fig.\ref{fig4}c and \ref{fig4}d are 1.5 T and 943 mT respectively , in good agreement with the measurements.

The current peak associated with tunneling into the excited (1,1) spin-triplet state, at $\epsilon = \Delta_{\rm ST}$, remains essentially unchanged up to $B\approx$ 5\,T.  This behavior is consistent with the expected B-evolution of the involved (1,1) and (0,2) triplet states, when considering only the Zeeman effect.

Figs. ~\ref{fig4}e and ~\ref{fig4}f present schematic energy diagrams of the $(1,1)$ and $(0,2)$ states at zero and finite $B$, respectively. 
The double dot system is effectively described by eight states in total: four $(1,1)$ states, including the lowest energy spin-singlet, $S_{11}$, and the three triplet states, $T^+_{11}$, $ T^0_{11}$, $T^-_{11}$; four $(0,2)$ states, including the spin singlet $S_{02}$ and three triplet states $T^+_{02}$, $ T^0_{02}$ and $T^-_{02}$. In the limit of weak inter-dot tunneling, all $(1,1)$ states are effectively degenerate at $B=0$ T. 
Two current peaks are thus expected: a first peak when $S_{11}$ and $S_{02}$ line up ($\epsilon=0$), and a second one when the degenerate $T_{11}$ states line up with the degenerate $T_{02}$ states ($\epsilon = \Delta_{\rm ST}$). At finite $B$, each spin triplet splits resulting in three non-degenerate levels separated by the Zeeman energy  $E_{Z}= g\mu_{B}B$. To first approximation we can reasonably assume $g$ to be equal to the bare electron g-factor in both QDs. In this case, triplet states with the same component that are aligned at $B=0$ (i.e. for $\epsilon = \Delta_{\rm ST}$) will remain aligned at all $B$ (see Fig.~\ref{fig4}f), which is consistent with our finding that the second peak at  $\epsilon = \Delta_{\rm ST}$ is essentially independent of $B$.   We note that, due to the extremely weak spin-orbit coupling in the silicon conduction band, tunneling between triplet states with different component is negligible. 

In summary we have studied spin-dependent transport in double QDs defined in a CMOS SOI nanowire transistor featuring two parallel top gates. At low temperature, devices tuned to the few electron regime exhibit Pauli spin blockade signatures. The revealed singlet-triplet splitting ranges from 0.3 to 1.3\,meV. The transitions which conserve spin are shown to be magnetic-field independent up to $B = 6$ T, which is expected for materials in which spin-orbit coupling is negligible and the main consequence of the field is to lift spin degeneracy through the Zeeman effect. Our results establish a first step towards a truly industrial silicon CMOS spin quantum bit.\\
\\
\textbf{Acknowledgments}
The research leading to these results has been supported by the European Union's through the research grants No. 323841, No. 610637, and No. 688539, as well as through the ERC grant No. 280043.

\bibliography{PauliV3.bib}

\end{document}